\documentclass[12pt]{article}
\usepackage{amsmath}
\usepackage{latexsym}
\usepackage{amsfonts}
\usepackage[normalem]{ulem}
\usepackage{array}
\usepackage{amssymb}
\usepackage[pdftex]{graphicx}
\usepackage[binary-units=true]{siunitx}
\DeclareSIUnit \dBm {dBm}
\DeclareSIUnit \dB {dB} 
\DeclareSIUnit \dBi {dBi} 
\DeclareSIUnit \Kbps {Kbps}
\DeclareSIUnit \Mbps {Mbps}
\DeclareSIUnit \Gbps {Gbps}
\DeclareSIUnit \kBps {kBps}
\DeclareSIUnit \MBps {MBps}
\DeclareSIUnit \GBps {GBps}
\DeclareSIUnit \GHz {GHz}
\usepackage{cite}
\usepackage{subfig}
\usepackage{wrapfig}
\usepackage{wasysym}
\usepackage{enumitem}
\usepackage{adjustbox}
\usepackage{ragged2e}
\usepackage[svgnames,table]{xcolor}
\usepackage{tikz}
\usepackage{longtable}
\usepackage{changepage}
\usepackage{setspace}
\usepackage{hhline}
\usepackage{multicol}
\usepackage{tabto}
\usepackage{float}
\usepackage{multirow}
\usepackage{makecell}
\usepackage{fancyhdr}
\usepackage{nicefrac}
\usepackage[toc,page]{appendix}
\usepackage[hidelinks]{hyperref}
\usepackage{tikz}
\def\checkmark{\tikz\fill[scale=0.4](0,.35) -- (.25,0) -- (1,.7) -- (.25,.15) -- cycle;} 
\usetikzlibrary{shapes.symbols,shapes.geometric,shadows,arrows.meta}
\tikzset{>={Latex[width=1.5mm,length=2mm]}}
\usepackage{flowchart}\usepackage[paperheight=11.0in,paperwidth=8.5in,left=1.0in,right=1.0in,top=1.0in,bottom=1.0in,headheight=1in]{geometry}
\usepackage[utf8]{inputenc}
\usepackage[T1]{fontenc}
\TabPositions{0.5in,1.0in,1.5in,2.0in,2.5in,3.0in,3.5in,4.0in,4.5in,5.0in,5.5in,6.0in,}
\setlistdepth{9}
\renewlist{enumerate}{enumerate}{9}
        \setlist[enumerate,1]{label=\arabic*)}
        \setlist[enumerate,2]{label=\alph*)}
        \setlist[enumerate,3]{label=(\roman*)}
        \setlist[enumerate,4]{label=(\arabic*)}
        \setlist[enumerate,5]{label=(\Alph*)}
        \setlist[enumerate,6]{label=(\Roman*)}
        \setlist[enumerate,7]{label=\arabic*}
        \setlist[enumerate,8]{label=\alph*}
        \setlist[enumerate,9]{label=\roman*}
\renewlist{itemize}{itemize}{9}
        \setlist[itemize]{label=$\cdot$}
        \setlist[itemize,1]{label=\textbullet}
        \setlist[itemize,2]{label=$\circ$}
        \setlist[itemize,3]{label=$\ast$}
        \setlist[itemize,4]{label=$\dagger$}
        \setlist[itemize,5]{label=$\triangleright$}
        \setlist[itemize,6]{label=$\bigstar$}
        \setlist[itemize,7]{label=$\blacklozenge$}
        \setlist[itemize,8]{label=$\prime$}
\renewcommand{\arraystretch}{1.3}

\begin{document}

\begin{table}[H]
             \centering
\begin{tabular}{p{2.96in}p{3.13in}}
\hline
\multicolumn{1}{|p{2.96in}}{\textbf{$\ast$ Title:\textit{\  }}} & 
\multicolumn{1}{|p{3.13in}|}{A Dataset of Full-Stack ITS-G5 DSRC Communications over Licensed and Unlicensed Bands Using a Large-Scale Urban Testbed} \\
\hhline{--}
\multicolumn{1}{|p{2.96in}}{\textbf{$\ast$ Authors:}} & 
\multicolumn{1}{|p{3.13in}|}{Andrea Tassi, Ioannis Mavromatis} \\
\hhline{--}
\multicolumn{1}{|p{2.96in}}{\textbf{$\ast$ Affiliations:\  }} & 
\multicolumn{1}{|p{3.13in}|}{Department of Electric and Electronic Engineering, University of Bristol, UK} \\
\hhline{--}
\multicolumn{1}{|p{2.96in}}{\textbf{$\ast$ Contact email: }} & 
\multicolumn{1}{|p{3.13in}|}{\{A.Tassi, Ioan.Mavromatis, R.J.Piechocki\} @bristol.ac.uk} \\
\hhline{--}
\multicolumn{1}{|p{2.96in}}{\textit{$\ast$ }\textbf{Co-authors}:\textit{ } \par } & 
\multicolumn{1}{|p{3.13in}|}{Robert Piechocki, Department of Electric and Electronic Engineering, University of Bristol, UK, and The Alan Turing Institute, London, NW1 2DB, UK, email: R.J.Piechocki@bristol.ac.uk} \\
\hhline{--}
\multicolumn{1}{|p{2.96in}}{\textbf{$\ast$ CATEGORY:}} & 
\multicolumn{1}{|p{3.13in}|}{Computer Science (Computer Networks and Communications)} \\
\hhline{--}

\end{tabular}
 \end{table}

\vspace{\baselineskip}

\vspace{\baselineskip}
\setlength{\parskip}{5.04pt}
\textbf{Data Article}\par

\textbf{Title}: A Dataset of Full-Stack ITS-G5 DSRC Communications over Licensed and Unlicensed Bands Using a Large-Scale Urban Testbed\par

\textbf{Authors}: Andrea Tassi$^*$, Ioannis Mavromatis$^*$, Robert Piechocki$^*$$^+$ \par

\textbf{Affiliations}: $^*$Department of Electric and Electronic Engineering, University of Bristol, UK, and \mbox{$^+$The} Alan Turing Institute, London, NW1 2DB, UK\par

\textbf{Contact email}: \{A.Tassi, Ioan.Mavromatis, R.J.Piechocki\}@bristol.ac.uk\par

\newpage\textbf{Abstract}\par
{ A dataset of measurements of ETSI ITS-G5 Dedicated Short Range Communications (DSRC) is presented. Our dataset consists of network interactions happening between two On-Board Units (OBUs) and four Road Side Units (RSUs). Each OBU was fitted onto a vehicle driven across the FLOURISH Test Track in Bristol, UK. Each RSU and OBU was equipped with two transceivers operating at different frequencies. During our experiments, each transceiver broadcasts Cooperative Awareness Messages (CAMs) over the licensed DSRC band, and over the unlicensed Industrial, Scientific, and Medical radio (ISM) bands $\SI{2.4}{\giga\hertz}$-$\SI{2.5}{\giga\hertz}$ and $\SI{5.725}{\giga\hertz}$-$\SI{5.875}{\giga\hertz}$.

Each transmitted and received CAM is logged along with its Received Signal Strength Indicator (RSSI) value and accurate positioning information. The Media Access Control layer (MAC) layer Packet Delivery Rates (PDRs) and RSSI values are also empirically calculated across the whole length of the track for any transceiver. The dataset can be used to derive realistic approximations of the PDR as a function of RSSI values under urban environments and for both the DSRC and ISM bands -- thus, the dataset is suitable to calibrate (simplified) physical layers of full-stack vehicular simulators where the MAC layer PDR is a direct function of the RSSI. The dataset is not intended to be used for signal propagation modelling.}

The dataset can be found at \url{https://doi.org/10.5523/bris.eupowp7h3jl525yxhm3521f57}, and it has been analyzed in the following paper: I. Mavromatis, A. Tassi, and R. J. Piechocki, ``Operating ITS-G5 DSRC over Unlicensed Bands: A City-Scale Performance Evaluation,'' \textit{IEEE PIMRC 2019}. [Online]. Available: \url{https://arxiv.org/abs/1904.00464}.

\vspace{\baselineskip}
\par

\newpage
\vspace*{-20mm}\textbf{Specifications Table }\par

\vspace{0mm}\begin{table}[H]
             \centering
\begin{tabular}{p{1.6in}p{4.5in}}
\hline
\multicolumn{1}{|p{1.6in}}{Subject area} & 
\multicolumn{1}{|p{4.5in}|}{Computer Science} \\
\hhline{--}
\multicolumn{1}{|p{1.6in}}{More specific subject area} & 
\multicolumn{1}{|p{4.5in}|}{Computer Networks and Communications} \\
\hhline{--}
\multicolumn{1}{|p{1.6in}}{Type of data} & 
\multicolumn{1}{|p{4.5in}|}{The network data exchanged by each RSU and OBU is provided as multiple PCAP traces~\cite{pcap} and CSV files. For each experimental session, each transceiver onboard each RSU and OBU is associated with a PCAP trace and CSV file. For each transceiver pertaining to each RSU and OBU, we also included a processed version of the data saved as MATLAB tables~\cite{mat}.} \\
\hhline{--}
\multicolumn{1}{|p{1.6in}}{How data was acquired} & 
\multicolumn{1}{|p{4.5in}|}{The Innovate UK-funded FLOURISH Test Track, which includes multiple RSUs and OBUs and spans a length of $\SI{5}{\kilo\metre}$ including key roads in the centre of Bristol, UK.} \\
\hhline{--}
\multicolumn{1}{|p{1.6in}}{Data format} & 
\multicolumn{1}{|p{4.5in}|}{Raw data format, as recorded by the RSUs and OBUs pertaining to the FLOURISH Test Track, in the form of PCAP traces and CSV files. The raw data are then filtered in order to remove not required information from the PCAP traces and the CSV files, such as the parts of an IEEE 802.11p frames encapsulating CAMs. Filtered data has been recorded as MATLAB tables.} \\
\hhline{--}
\multicolumn{1}{|p{1.6in}}{Parameters for data collection} & 
\multicolumn{1}{|p{4.5in}|}{The data has been recorded by employing two vehicles driving around the whole length of the FLOURISH Test Track. Since the data set has been recorded across multiple days, we relied on different drivers driving for $\sim\SI{4}{\hour}$ per day (namely, $\sim\SI{2}{\hour}$ in the morning and $\sim\SI{2}{\hour}$ in the afternoon). Furthermore, drivers experienced different traffic conditions in the city -- thus the time needed to drive the whole length of the track depended on the time of the day.
} \\
\hhline{--}
\multicolumn{1}{|p{1.6in}}{Description of data collection} & 
\multicolumn{1}{|p{4.5in}|}{
Eight data recording sessions of $\sim\SI{2}{\hour}$ each, where one vehicle was driven in a clockwise and the other in an anticlockwise fashion across the whole length of the  FLOURISH Test Track.

Two transmission frequencies were tested during each data recording session (one per transceiver).} \\
\hhline{--}
\multicolumn{1}{|p{1.6in}}{Data source location} & 
\multicolumn{1}{|p{4.5in}|}{A typical urban setting in Bristol, UK, comprising the following roads: Marlborough Street, Upper Maudlin Street, Park Row, Woodland Road and Queens Road.} \\
\hhline{--}
\multicolumn{1}{|p{1.6in}}{Data accessibility} & 
\multicolumn{1}{|p{4.5in}|}{Data is available online at the link below: \newline \url{https://doi.org/10.5523/bris.eupowp7h3jl525yxhm3521f57}} \\
\hhline{--}
\multicolumn{1}{|p{1.6in}}{Related research article} & 
\multicolumn{1}{|p{4.5in}|}{I. Mavromatis, A. Tassi and R. J. Piechocki, ``Operating ITS-G5 DSRC over Unlicensed Bands: A City-Scale Performance Evaluation,'' \textit{IEEE PIMRC 2019}. [Online]. Available: \url{https://arxiv.org/abs/1904.00464}.} \\
\hhline{--}

\end{tabular}
 \end{table}


\newpage
\textbf{Value of the Data}\par
\begin{itemize}
\item 
{Our dataset consists of network interactions recorded among RSUs and OBUs in an urban environment in both a Vehicle-to-Infrastructure (V2I) and Vehicle-to-Vehicle (V2V) fashion. Network interactions account for CAMs exchanged according to the standard ITS-G5 DSRC, operated over the licensed DSRC band and unlicensed ISM bands $\SI{2.4}{\giga\hertz}$-$\SI{2.5}{\giga\hertz}$ and $\SI{5.725}{\giga\hertz}$-$\SI{5.875}{\giga\hertz}$. Along with the dataset we also provided the necessary data processing tools to estimate the MAC layer PDR as a function of RSSI values. The empirical PDR value can be directly integrated into (simplified) physical layers that are widely adopted in full-stack network simulators. It is beyond the scope of our experiments to perform a measurement campaign suitable for deriving accurate propagation models.}

\item 
{By building upon the empirical PDR values (expressed as a function of the RSSI values), experts from the Wireless Communications and Networks communities can directly assess the impact of operating an ITS-G5 DSRC system over both licensed and unlicensed frequency bands, being respectively the DSRC and the ISM bands. For instance, this will facilitate further research on the provision of safety-critical and non-safety critical automotive services over the DSRC band, the ISM bands, or both at the same time.}

\item 
{Licensed DSRC band is expected to be (at least partially) allocated to Cellular-V2X (C-V2X) systems. As such, the empirical PDR values (expressed as a function of RSSI values) derived from this dataset can be directly compared against empirical PDR values measured for C-V2X systems, in urban environments. Furthermore, a similar performance investigation can be carried out by comparing the empirical PDR values of C-V2X against ITS-G5 DSRC operated over the unlicensed ISM bands we considered in our experiments.}

\item 
{Since our dataset has been recorded employing vehicles driving in a real urban environment; it can also be used by cyber-security experts. A potential usage is to establish what the ``standard behavior'' of a connected vehicle is when observed from the communication perspective -- thus making it possible to train anomaly detection systems to identify potential threats~\cite{Cyber2019}. For instance, a vehicle can maliciously broadcast CAMs where its location has been altered. However, any RSU receiving malicious CAMs will be able to measure their RSSI values. Since the RSSI value is a function of the environment and the distance between the transmitter and the receiver, an anomaly detection system can verify if a measured RSSI value is comparable with the expected value given the positioning information encapsulated into each CAM.}
\end{itemize}

\vspace{\baselineskip}
\textbf{Data}\par
The raw data associated with each transceiver (namely, PCAP traces~\cite{pcap} and CSV files) installed on each RSU and OBU are organized as follows. During each day of trial, our dataset includes one or two sequences of raw data files for each transceiver. That depends on if the transceiver is installed on either an RSU or an OBU. In particular, raw data files associated with each RSU collates two experimental sessions (namely, morning and afternoon) per-file, while raw data files pertaining to each OBU always refer to a single experimental session.

As for the PCAP traces, each of them records the whole of the network interaction where each transceiver is involved in. In particular, each received IEEE 802.11p frame is recorded into a PCAP trace with its corresponding RSSI value. Our PCAP traces can be accessed using our modified version of the tool Tcpdump~\cite{tcpdump}. Furthermore, for each transceiver, key fields of each of the transmitted and received CAMs are recorded in a tabular format in the same CSV file. Since, the FLOURISH Testing Track logs both the transmitted and received CAMs in the same raw data file, for the sake of data readability and debugging purposes, the labels of each column are replicated in each data row, i.e., each data row follows the format:
\begin{center}
    {\tt Column\_Label\_0;Value\_0;Column\_Label\_1;Value\_1; $\ldots$ Column\_Label\_n;Value\_n}
\end{center}

For the sake of reducing the overall size of the dataset and hence, making it possible to manipulate the dataset with a smaller memory footprint, the raw data has been filtered into MATLAB data files~\cite{mat}. In particular, for each set of CSV file and PCAP trace associated to a given transceiver, we generated the corresponding MATLAB tables for the transmitted/received CAMs, and one MATLAB table per-PCAP trace referring to the received CAMs. Further details about the labels of the columns in the CSV raw data files and MATLAB tables are provided in Table~\ref{tab.1}.

{The MATLAB framework designed for the above data manipulation and filtering can be found online in~\cite{matScripts}. Our framework provides also tools to investigate the Key Performance Indicators (KPIs) shown in Figs.~\ref{fig.1}-\ref{fig.8}. These figures will be further discussed in the next section.}

\begin{table}[t]
\centering
\caption{Definition of the labels of each column of the CSV raw data files and MATLAB tables in the dataset.}
    \label{tab.1}
\scriptsize
\begin{tabular}{|c|l|p{80mm}|c|c|}
\hline
\multicolumn{2}{|c|}{\multirow{2}{*}{\textbf{Field Name}}} & \multicolumn{1}{c|}{\multirow{2}{*}{\textbf{Definition}}}                       & \multirow{2}{*}{\textbf{TX Entry}} & \multirow{2}{*}{\textbf{RX Entry}} \\
\multicolumn{2}{|c|}{}                                     & \multicolumn{1}{c|}{}                                                           &                                    &                                    \\ \hline
\multicolumn{2}{|c|}{\tt TX-REQ-CAM}                           & Transmitted CAM                                                                 & \checkmark      &                                    \\ \hline
\multicolumn{2}{|c|}{\tt RxMAC}                                & MAC address of the transmitter                                                  &                                    & \checkmark      \\ \hline
\multicolumn{2}{|c|}{\tt RX-REQ-CAM}                           & Received CAM                                                                    &                                    & \checkmark      \\ \hline
\multicolumn{2}{|c|}{\tt Protocol}                             & ITS-G5 protocol ID~\cite{etsiCam}                         & \checkmark      &                                    \\ \hline
\multicolumn{2}{|c|}{\tt Validation}                           & ITS-G5 validation ID~\cite{etsiCam}                       &                                    & \checkmark      \\ \hline
\multicolumn{2}{|c|}{\tt StationID}                            & ID of the transmitter/receiver                                                  & \checkmark      & \checkmark      \\ \hline
\multicolumn{2}{|c|}{\tt GenDeltaTime}                         & The remainder of $Timestamp/65546$~\cite{etsiCam}         & \checkmark      & \checkmark      \\ \hline
\multicolumn{2}{|c|}{\tt SeqNum}                               & Sequence number of the transmitted/received CAM~\cite{2019arXiv190301377T} & \checkmark      & \checkmark      \\ \hline
\multicolumn{2}{|c|}{\tt GpsLon}                               & Longitude of the current position of the transmitter/receiver                   & \checkmark      & \checkmark      \\ \hline
\multicolumn{2}{|c|}{\tt GpsLat}                               & Latitude of the current position of the transmitter/receiver                    & \checkmark      & \checkmark      \\ \hline
\multicolumn{2}{|c|}{\tt CamLon}                               & Longitude of the position of the transmitter/receiver as encapsulated in a CAM  & \checkmark      & \checkmark      \\ \hline
\multicolumn{2}{|c|}{\tt CamLat}                               & Latitude of the position of the transmitter/receiver as encapsulated in a CAM   & \checkmark      & \checkmark      \\ \hline
\multicolumn{2}{|c|}{\tt SpeedConf}                            & Encoding configuration for the speed~\cite{etsiCam}            &                                    & \checkmark      \\ \hline
\multicolumn{2}{|c|}{\tt VehHeading}                           & Heading of the transmitter                                                      &                                    & \checkmark      \\ \hline
\multicolumn{2}{|c|}{\tt GpsSpeed}                             & Speed of the transmitter/receiver                                               & \checkmark      & \checkmark      \\ \hline
\multicolumn{2}{|c|}{\tt CamSpeed}                             & Speed of the transmitter as encapsulated in a CAM                               & \checkmark      & \checkmark      \\ \hline
\multicolumn{2}{|c|}{\tt Timestamp}                            & Timestamp at the moment of creation of a CAM                                    & \checkmark      & \checkmark      \\ \hline
\multicolumn{2}{|c|}{\tt CamLength}                            & Byte-length of a CAM                                                            & \checkmark      &                                    \\ \hline
\multicolumn{2}{|c|}{\tt RSSI}                            & The RSSI value of a received CAM                                                            &       &     \checkmark                               \\ \hline
\end{tabular}
\end{table}

\vspace{\baselineskip}

\vspace{\baselineskip}
\newpage\textbf{Experimental Design, Materials, and Methods}\par

\setlength{\parskip}{9.96pt}

\begin{figure}[t]     
\centering
    \includegraphics[width=1\columnwidth]{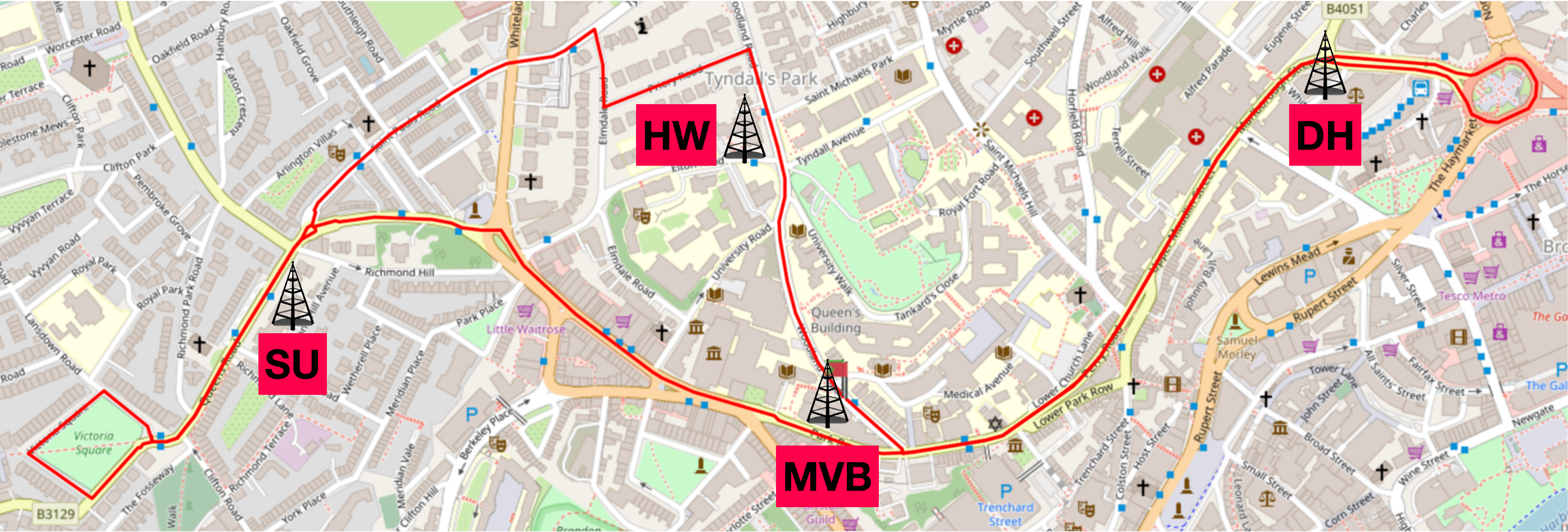}
    \caption{The Flourish Testing Track and the locations of the four RSUs.}
    \label{fig.0}
\end{figure}

The four RSUs providing network coverage across the $\SI{5}{\kilo\metre}$ length of the FLOURISH Testing Track (see Fig.~\ref{fig.0}) are fitted outside premises of the University of Bristol. Each site, along with its ID and characteristics, is listed below:
\begin{itemize} 
    \item MVB -- \emph{Merchant Ventures Building, BS8 1UB, UK}: RSU installed at a height of $\sim$\SI{8}{\meter} opposite to a T-junction. 
    \item DH -- \emph{Dorothy Hodgkin Building, BS1 3NY, UK}: RSU located at $\sim$\SI{12}{\meter} from the ground beside a curvy road. 
    \item HW -- \emph{Hawthorns Building, BS8 1UQ, UK}: RSU mounted at $\sim$\SI{5}{\meter} from street-level, some foliage on site.
    \item SU -- \emph{Students Union Building, BS8 1LN, UK}: RSU fitted at $\sim$\SI{25}{\meter} from the ground beside a straight stretch of road.
\end{itemize}
All aforementioned sites host an identical RSU operating two transceivers at the same time. Since the first transceiver has a maximum transmission power of $\SI{25}{\dBm}$ and the second can reach up to $\SI{29}{\dBm}$, in our dataset, we regard the first one as the Low Power (LP) transceiver and the second one as the High Power (HP) transceiver.

During each experimental session, two vehicles have also been used, with the only exception of the afternoon experimental session held on $15^{\text{th}}$ February 2019 when only one vehicle had been used. When two vehicles were present, one was driven in a clockwise direction while the other in an anticlockwise fashion. Each vehicle was equipped with an OBU operating the same set of transceivers that we employed in each RSU. Further details about both the hardware and software pertaining to the RSUs and OBUs can be found in~\cite{PIMRC}. { During each day of trials, the LP and HP transceivers onboard each RSU and OBU have been operated on $\SI{10}{\mega\hertz}$-width channels. The center frequencies chosen for each day, the considered transmission power and Modulation and Coding Scheme (MCS) are listed in Table~\ref{table:freq}. Since the data traffic exchanged between the OBUs and RSUs was entirely made of BTP and CAM messages, we adopted the MCS QPSK-1/2 as imposed by the ITS-G5 DSRC standard~\cite{etsiMCS}.}

\begin{table}[t]
\renewcommand{\arraystretch}{1.07}
\centering
\scriptsize
\caption{The center frequency, transmission power and MCS used throughout the four days of trials.}
    \begin{tabular}{|c|c|c|c|c|c|c|}
    \hline\textbf{Date} & \textbf{HP Transceivers} & \textbf{LP Transceivers} & \textbf{Band} & {\textbf{HP Power}} & {\textbf{LP Power}} & {\textbf{MCS}~\cite{etsiMCS}}  \\ \hline
    $1^{\text{st}}$ Day -- 11/02/2019 & \SI{5.900}{\GHz} & \SI{5.890}{\GHz} & DSRC & \multirow{4}{*}{\SI{29}{\dBm}} & \multirow{4}{*}{\SI{25}{\dBm}}  & \multirow{4}{*}{QPSK \nicefrac{1}{2}} \\
    $2^{\text{nd}}$ Day -- 13/02/2019 & \SI{5.200}{\GHz} & \SI{2.437}{\GHz} & ISM & & & \\
    $3^{\text{rd}}$ Day -- 15/02/2019 & \SI{5.180}{\GHz} & \SI{2.412}{\GHz} & ISM & & & \\
    $4^{\text{th}}$ Day -- 18/02/2019 & \SI{5.320}{\GHz} & \SI{2.462}{\GHz} & ISM & & & \\ \hline
    \end{tabular}
\label{table:freq}
\end{table}

\begin{figure}[t]
\centering
    \subfloat[{Format of the adopted CAM.}]{\includegraphics[width=0.19\columnwidth]{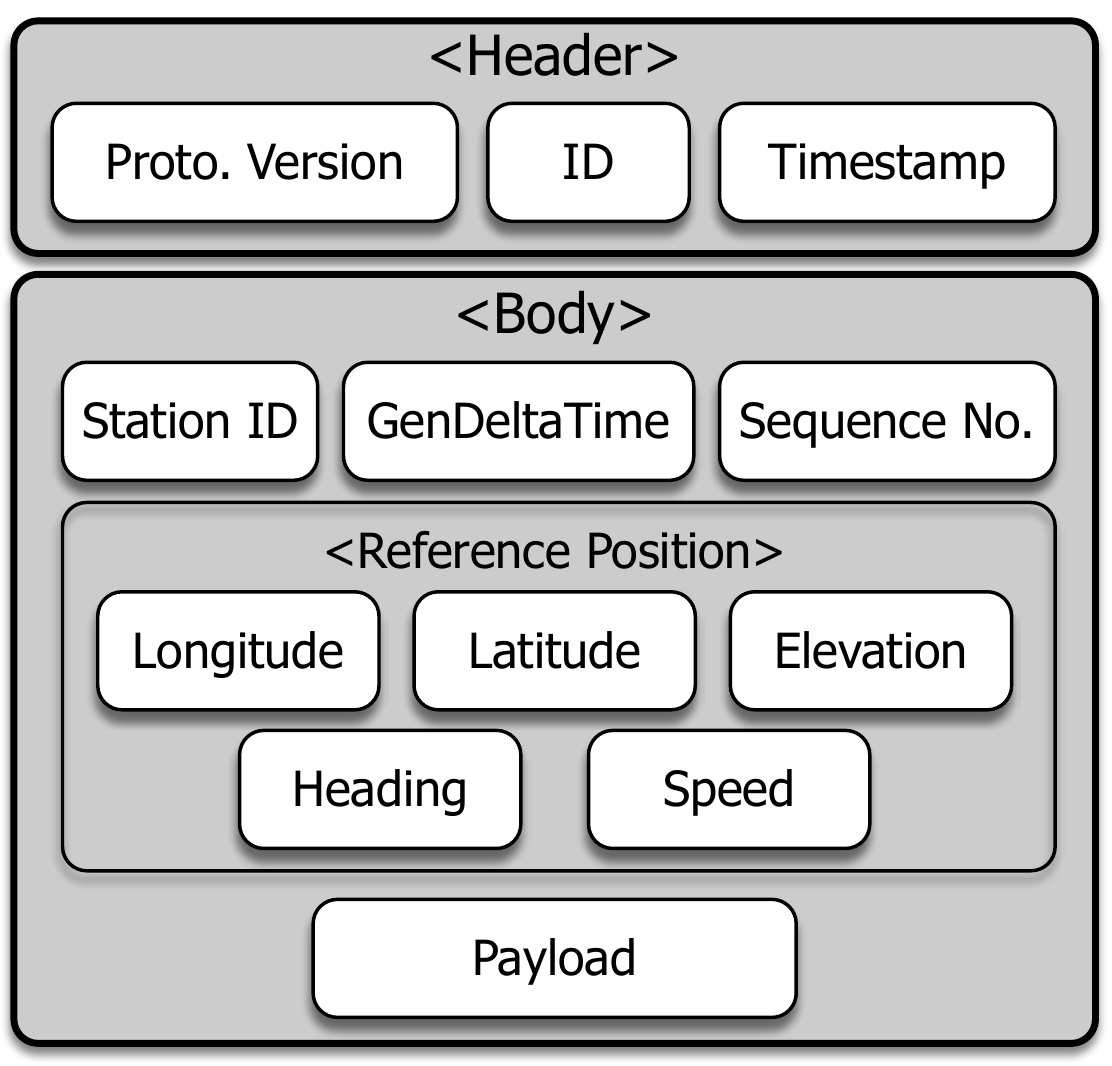}}
    \:
    \subfloat[{The flowchart showing the decisions for recording a CAM within the dataset.}]{\includegraphics[width=0.79\columnwidth]{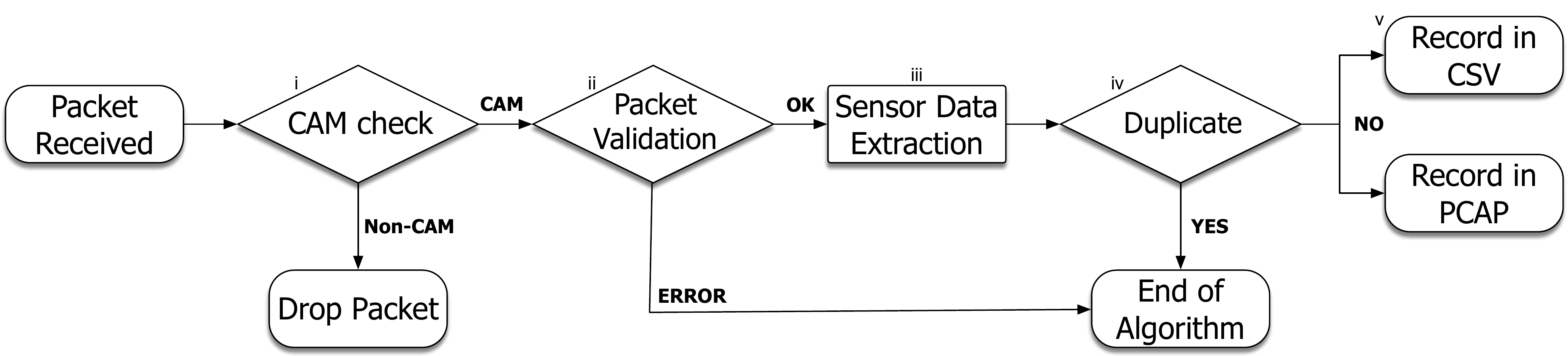}}
    \caption{Considered CAM and flowchart of the data logging process. When a packet is received in the TX or the RX queue, it is checked whether it is CAM message at first. Later, the validity of the packet is specified before being recorded in our dataset.}
    \label{fig.CAM}
\end{figure}

{ All the RSUs and OBUs communicate using ETSI's ITS-G5 standard~\cite{etsiStandard} and broadcast a CAM every $\SI{10}{\milli\second}$. Our setup is built upon the IEEE 802.11p DSRC physical layer as described before. On top of that, a MAC and Logical Link Control (LLC) layers act as the adaptation between the physical and the upper layers. Later, the Basic Transport Protocol (BTP) operating on top of the MAC and the LLC, provides an end-to-end connection-less transport service in an ad-hoc fashion. The BTP is also responsible for multiplexing/de-multiplexing packets originating from/to the Facilities layer~\cite{etsiStandard}. The exchanged CAMs are generated at the level of the Facilities layer. During our trials, we only referred to CAMs carrying positioning information (namely, position, speed and heading) pertaining to each OBUs and RSUs -- obviously, in the case of the RSUs, both speed and heading is set to zero. For our research activities, we extended the standard CAM format described in~\cite{etsiCam} by adding two extra fields, namely, the \emph{Timestamp} and the \emph{Sequence number}. The usage of these fields has been described in~\cite{PIMRC}, and it allowed us to reconcile the logged data during the data processing phase. All the fields encapsulated within the exchanged CAMs can be seen in Fig.~\ref{fig.CAM}(a).

Each transmitted and received CAM has been recorded both in the CSV and PCAP traces. When a packet appears in the transmission or reception queue of a transceiver, a decision is made whether it is going to be recorded or not. The flowchart describing that, can be seen in Fig.~\ref{fig.CAM}(b). For the sake of limiting the size of the dataset, only the CAMs exchanged were recorded in the traces (Fig.~\ref{fig.CAM}(b)-i). The rest of the wireless networks interactions within our nodes are discarded. The validity of CAMs within the queues is then checked (Fig.~\ref{fig.CAM}(b)-ii) by evaluating the length of the packet, and the expected fields based on the legacy CAM structure. For each valid CAM, the positioning information is extracted (Fig.~\ref{fig.CAM}(b)-iii). Subsequently, duplicated packets are then removed from the transmission/reception queue (Fig.~\ref{fig.CAM}(b)-iv). Whenever, the CAM is not valid or duplicates, the messaged is not recorded in the log files. Otherwise, the message is logged (Fig.~\ref{fig.CAM}(b)-v).}

In particular, for what concerns an RSU, raw data recorded during a specific day of trial and pertaining to a specific transceiver, is recorded under the following filename and directory naming convention:
\begin{center}
    {\tt ./rawData/ddmmyyyy/RSUs-ddmmyyyy/log\_XX\_YY.txt}
\end{center}
and
\begin{center}
    {\tt ./rawData/ddmmyyyy/RSUs-ddmmyyyy/log\_XX\_TCP\_YY.pcap}
\end{center}
where ``ddmmyyyy'' signifies the date when the experiment took place, ``XX'' can either be ``HP'' or ``LP'' depending on the transceiver under consideration, and ``YY'' refers to one of the four RSUs (namely, ``MVB'', ``DH'', ``HW'' or ``SU''). Thus, for instance, files ``./rawData/11022019/RSUs-11022019/log\_HP\_MVB.txt'' and ``./rawData/11022019/RSUs-11022019/log\_HP\_TCP\_MVB.txt'' refer to the HP transceiver of the RSU located on the MVB site during the $11^\text{th}$ February 2019. A similar conversion has been adopted for recording raw data associated with the OBUs, which is defined as follows:
\begin{center}
    {\tt ./rawData/ddmmyyyy/OBU-ZZ-ddmmyyyy-TT/log\_XX.txt}
\end{center}
and
\begin{center}
    {\tt ./rawData/ddmmyyyy/OBU-ZZ-ddmmyyyy-TT/log\_XX\_tcpd.pcap}
\end{center}
where ``ZZ'' can be either equal to ``00'' or ``01'' depending of the fact the data refers to the first or the second OBU, respectively. Then, if data refers to an experimental session that took place in the morning then ``TT'' is equal to ``Morning'', otherwise, ``TT'' is set equal to ``Afternoon''. For instance, files ``./rawData/11022019/OBU-00-11022019-Morning/log\_HP.txt'' and ``./rawData/11022019/OBU-00-11022019-Morning/log\_HP\_tcpd.pcap'' refer to the experimental session recorded in the morning of $11^{\text{th}}$ February 2019 and referring to the HP transceiver of the first OBU.

The raw data part of our dataset has been filtered by means of our MATLAB framework~\cite{matScripts}. The output of the data filtering operations has been made available under the directories ``importedData/ddmmyyyy'', where files follow the naming convention summarized in Table~\ref{tab.names}.

\begin{table}[t]
\renewcommand{\arraystretch}{1.07}
\centering
\caption{Naming conventions and per-file content description for the filtered part of our dataset.}
\scriptsize
    \begin{tabular}{|c|p{80mm}|}
    \hline\textbf{File Name} & \textbf{Content} \\ \hline
    {\tt rsu\_YY\_TxCAM\_XX.mat} &  Data from CAMs transmitted by the RSU at site ``YY'' (namely, ``MVB'', ``DH'', ``HW'' or ``SU''), from the transceiver ``XX'' (namely, ``LP'', or ``HP'') \\ \hline
    {\tt rsu\_YY\_RxCAM\_XX.mat} &  Data from CAMs received by the RSU at site ``YY'' (namely, ``MVB'', ``DH'', ``HW'' or ``SU''), from the transceiver ``XX'' (namely, ``LP'', or ``HP'') \\ \hline
    {\tt rsu\_YY\_tcpDump\_XX.mat} &  Data originating from the PCAP traces associated with the RSU at site ``YY'' (namely, ``MVB'', ``DH'', ``HW'' or ``SU''), from the transceiver ``XX'' (namely, ``LP'', or ``HP'') \\ \hline
    {\tt vehNo\_ZZ\_TxCAM\_XX\_UU.mat} &  Data from CAMs transmitted by the OBU at site ``ZZ'' (namely, ``00'' or ``01''), from the transceiver ``XX'' (namely, ``LP'', or ``HP''), during the ``UU'' session (namely, morning ``mr'' or afternoon ``af'')\\ \hline
    {\tt vehNo\_ZZ\_RxCAM\_XX\_UU.mat} &  Data from CAMs received by the OBU at site ``ZZ'' (namely, ``00'' or ``01''), from the transceiver ``XX'' (namely, ``LP'', or ``HP''), during the ``UU'' session (namely, morning ``mr'' or afternoon ``af'')\\ \hline
    {\tt vehNo\_ZZ\_tcpDump\_XX\_UU.mat} &  Data originating from the PCAP traces associated with the OBU at site ``ZZ'' (namely, ``00'' or ``01''), from the transceiver ``XX'' (namely, ``LP'', or ``HP''), during the ``UU'' session (namely, morning ``mr'' or afternoon ``af'') \\ \hline
    \end{tabular}
\label{tab.names}
\end{table}

{ Our MATLAB framework~\cite{matScripts} can also be used to investigate the following  Key Performance Indicators (KPIs). For simplicity, a representative subset of figures was chosen for each KPI:
\begin{itemize}
    \item Heatmaps for the PDR values associated with the HP/LP transceivers across the roads forming the FLOURISH Testing Track. A sample of the sixteen resulting figures is shown in Figs.~\ref{fig.1}-\ref{fig.4}. A CAM is considered as successfully delivered, when all the encapsulated sensor information can be successfully extracted (as shown in Fig.~\ref{fig.CAM}(b)).
    \item Awareness horizons for both HP and LP transceivers showing the PDR as a function of the distance to a given RSU. A sample of the sixteen resulting figures is shown in Figs.~\ref{fig.5}-\ref{fig.6}. The successful CAM reception is considered as in the heatmaps KPI.
    \item RSSI value for both HP and LP transceivers as a function of the distance to a given RSU. A sample of the thirty-two resulting figures is shown in Figs.~\ref{fig.7}-\ref{fig.8}.
\end{itemize}

Our framework can be used to generate the entire set of figures for all the above-mentioned KPIs. It can also be further extended to accommodate different or more specific KPIs pertaining to the network interactions at specific road junctions at specific time of the day, or the network interactions among moving vehicles. As for the latter, for convenience, our MATLAB framework~\cite{matScripts} includes a complementary script (namely, ``v2vInteractions/v2vTimings.m'') to locate the the timestamps when each vehicle received a CAM from another one, and saves them into MATLAB table.}

{Our MATLAB framework also gives the reader the option of filtering the entire dataset in order to retrieve all the network interactions between the pair of OBUs considered in our experiments. In particular, when the script ``v2vInteractions/v2vInteractions.m'' is executed, for each day and transceiver, it produces a table-formatted MAT file defined by the columns summarized in Table~\ref{tab.1} where each row corresponds to a transmitted/received CAM. For the fields that exist on both the TX and the RX sides (for e.g. the ``GpsLon''), we distinguish their values in the table by adding either ``-RX'' or ``-TX'' at the column name\footnote{{For simplicity we focused on the position of the vehicles (both the one from the GPS receiver and the one encapsulated in each CAM), the TX timestamp and the MAC addresses of the OBUs. The rest (such as the ``VehHeading'' for example), can be easily added by extending the ``v2vInteractions/v2vInteractions.m'' script and adding the new columns into the table.}} (for e.g. ``GpsLon-TX'' and ``GpsLon-RX''). Similarly, the reader can develop a filtering script to retrieve the network interactions pertaining to a given OBU and RSU. The starting point can be any of the existing scripts under the folder ``scriptResults'' folder that can be modified as follows:
\begin{enumerate}
    \item Modifying the two main loops of the script, namely, the loops associated with the day and the transmission power. The user can then choose to evaluate the results for a specific day or TX power. Based on that, our framework will load only the network interactions related to the chosen day and TX power.
    \item Modifying the internal loops of our script, namely, the ones related with the OBUs and the RSUs. One or more specific pairs of OBU/RSU can be chosen to be processed.
    \item Similarly to the ``v2vInteractions.m'' script, the unique sequence number for the transmitted packets per session, as well as the MAC address from the transmitter can be parsed at first.
    \item Using the above information, it can be identified whether a packet is successfully delivered or not, i.e., checking that a packet with the given MAC address and sequence number exists in the list of received packets.
    \item Having this pairwise correlation between transmitted and received packets, a table can be later generated incorporating all the given interactions between a set of devices (as in ``v2vInteractions.m'').
    \item The above steps can be replicated for the requested number of OBU-RSU pairs generating a list of all the desired network interactions.
\end{enumerate}}


\begin{figure}[t]
\centering
\subfloat[LP Transceiver]{
    \includegraphics[width=1\columnwidth]{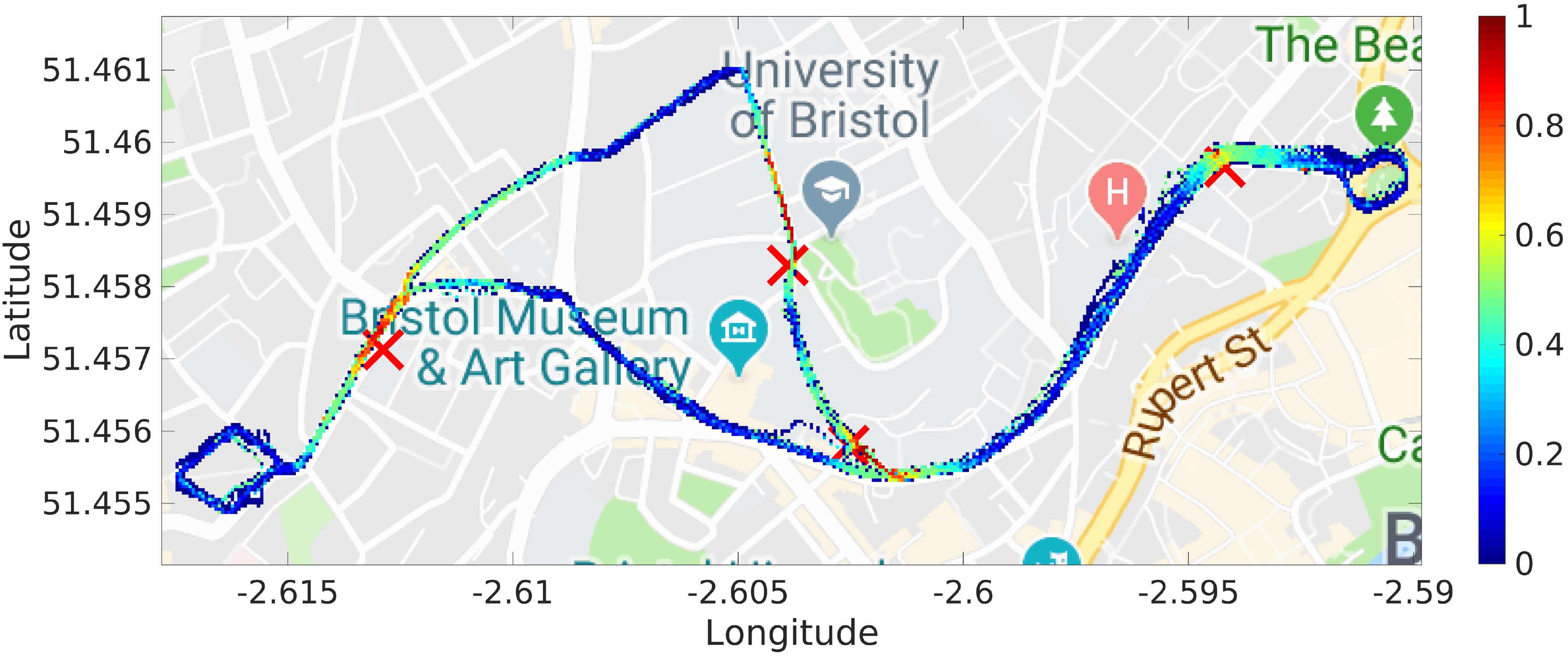}
}
\\
\subfloat[HP Transceiver]{
    \includegraphics[width=1\columnwidth]{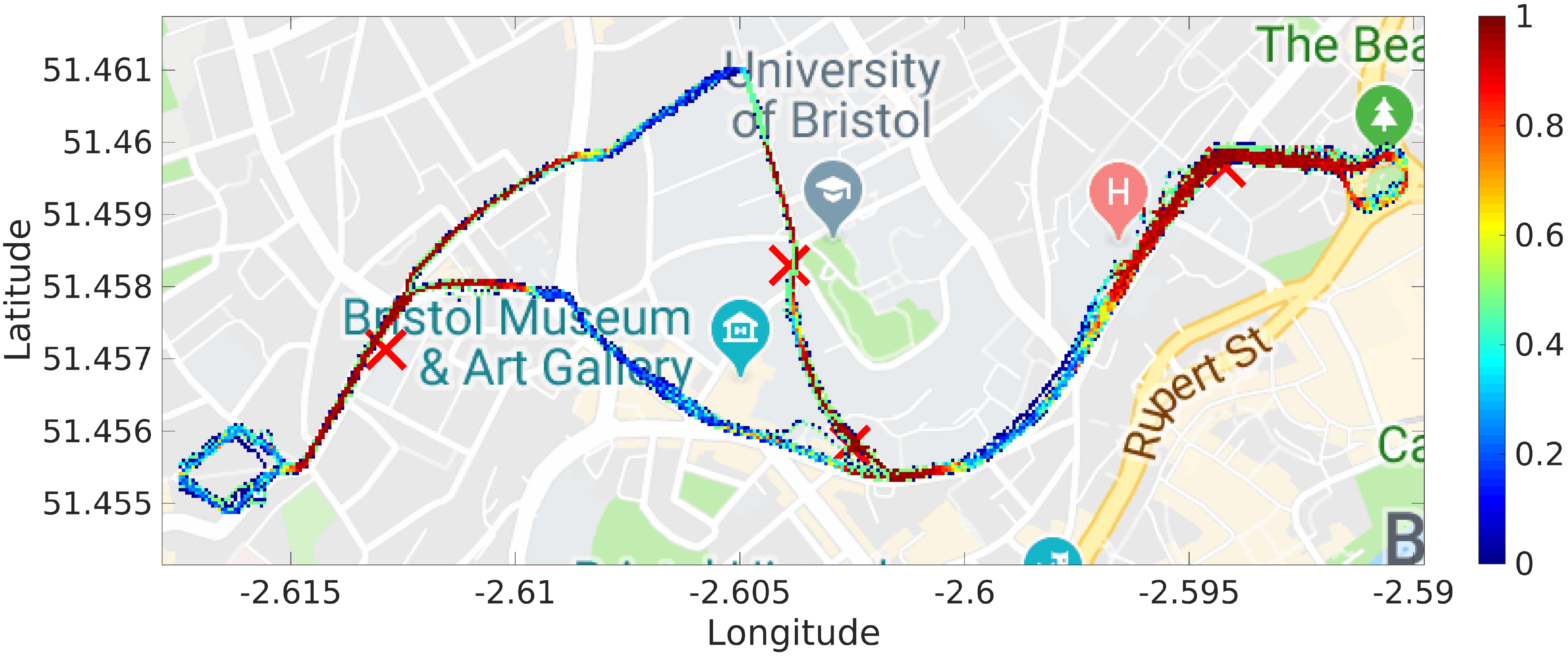}
}
\caption{Heatmaps for the PDR associated with the first day of trials (DSRC band).}
\label{fig.1}
\end{figure}

\begin{figure}[t]
\centering
\subfloat[LP Transceiver]{
    \includegraphics[width=1\columnwidth]{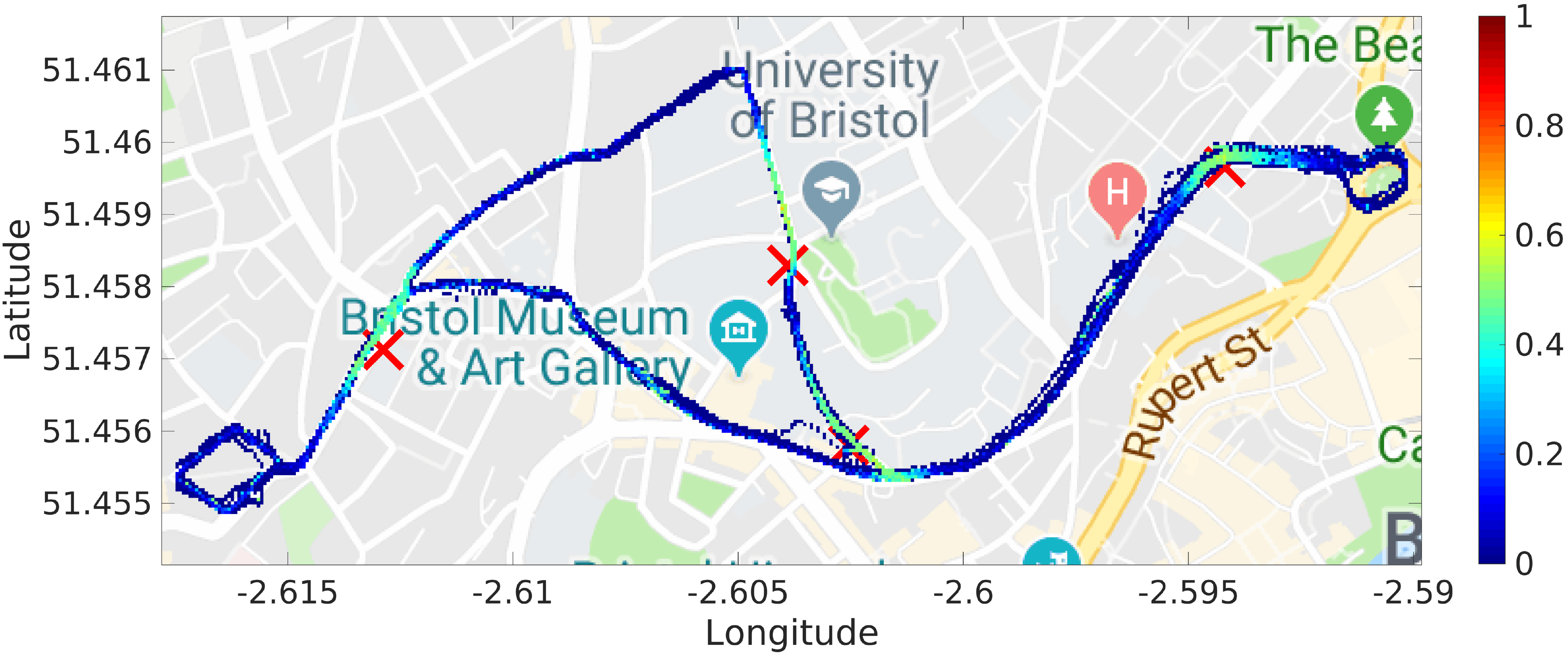}
}
\\
\subfloat[HP Transceiver]{
    \includegraphics[width=1\columnwidth]{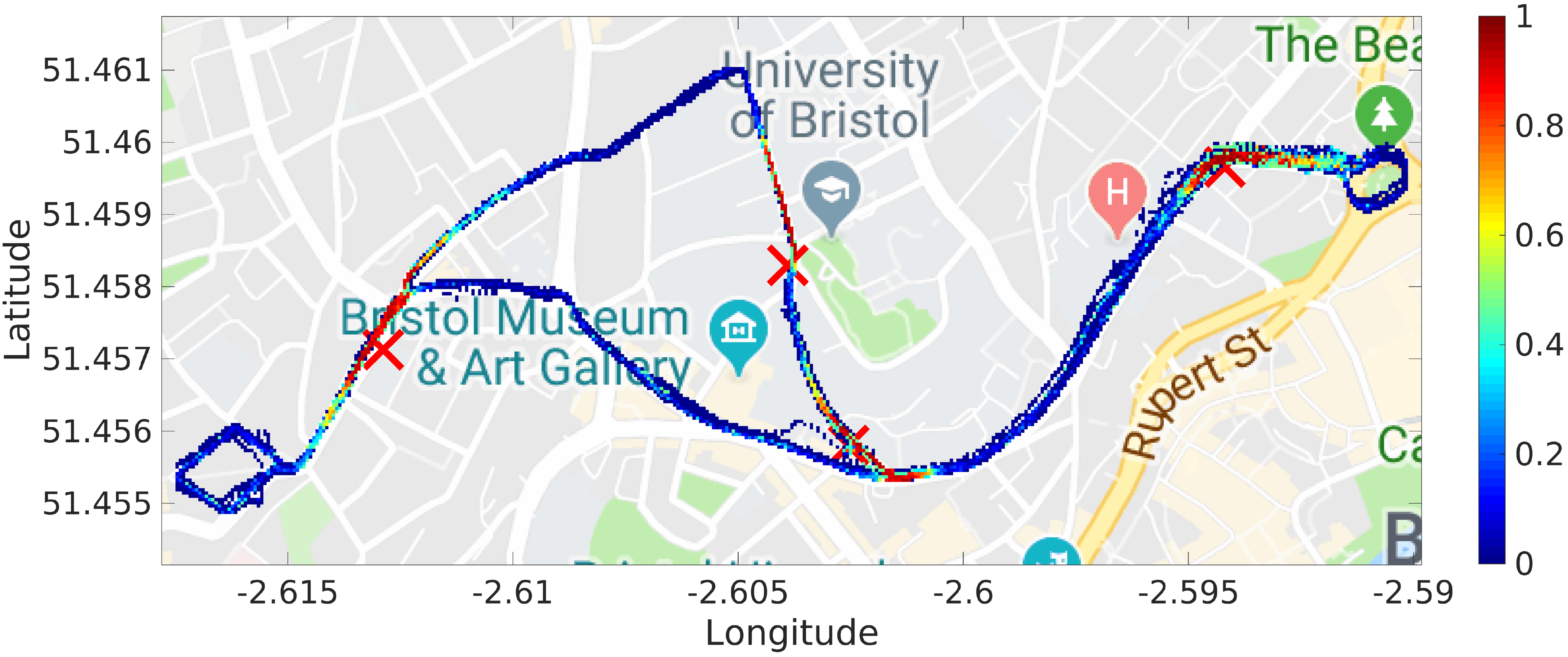}
}
\caption{Heatmaps for the PDR associated with the third day of trials (ISM band).}
\label{fig.2}
\end{figure}

\begin{figure}[t]
\centering
\subfloat[LP Transceiver]{
    \includegraphics[width=1\columnwidth]{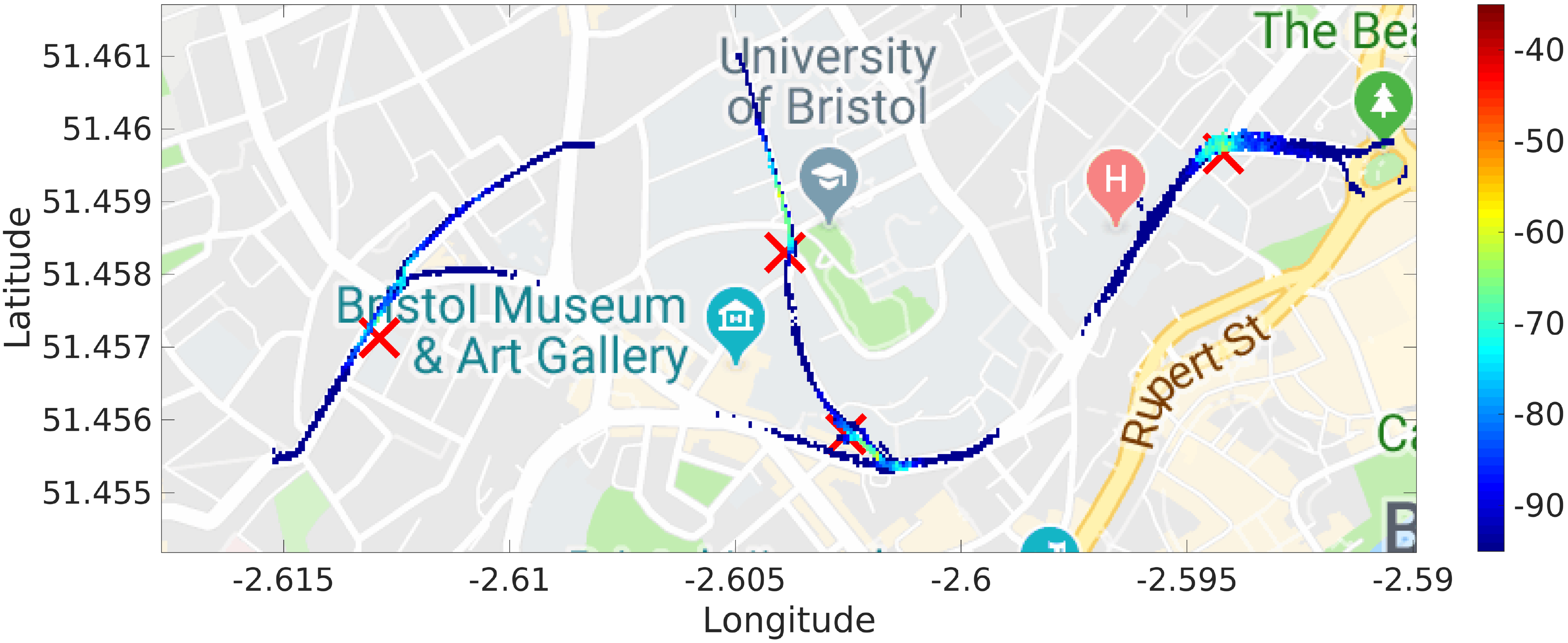}
}
\\
\subfloat[HP Transceiver]{
    \includegraphics[width=1\columnwidth]{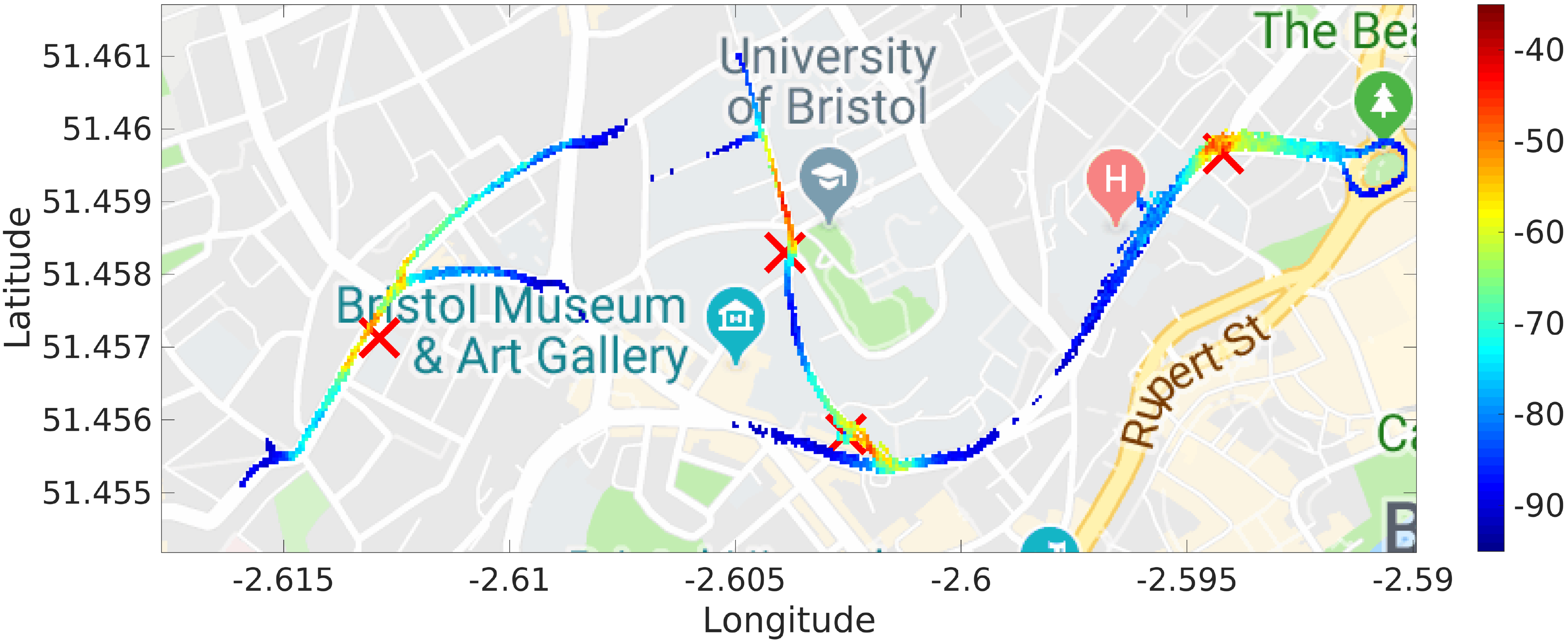}
}
\caption{Heatmaps for the RSSI values (expressed in $\SI{}{\dBm}$) associated with the first day of trials (DSRC band).}
\label{fig.3}
\end{figure}

\begin{figure}[t]
\centering
\subfloat[LP Transceiver]{
    \includegraphics[width=1\columnwidth]{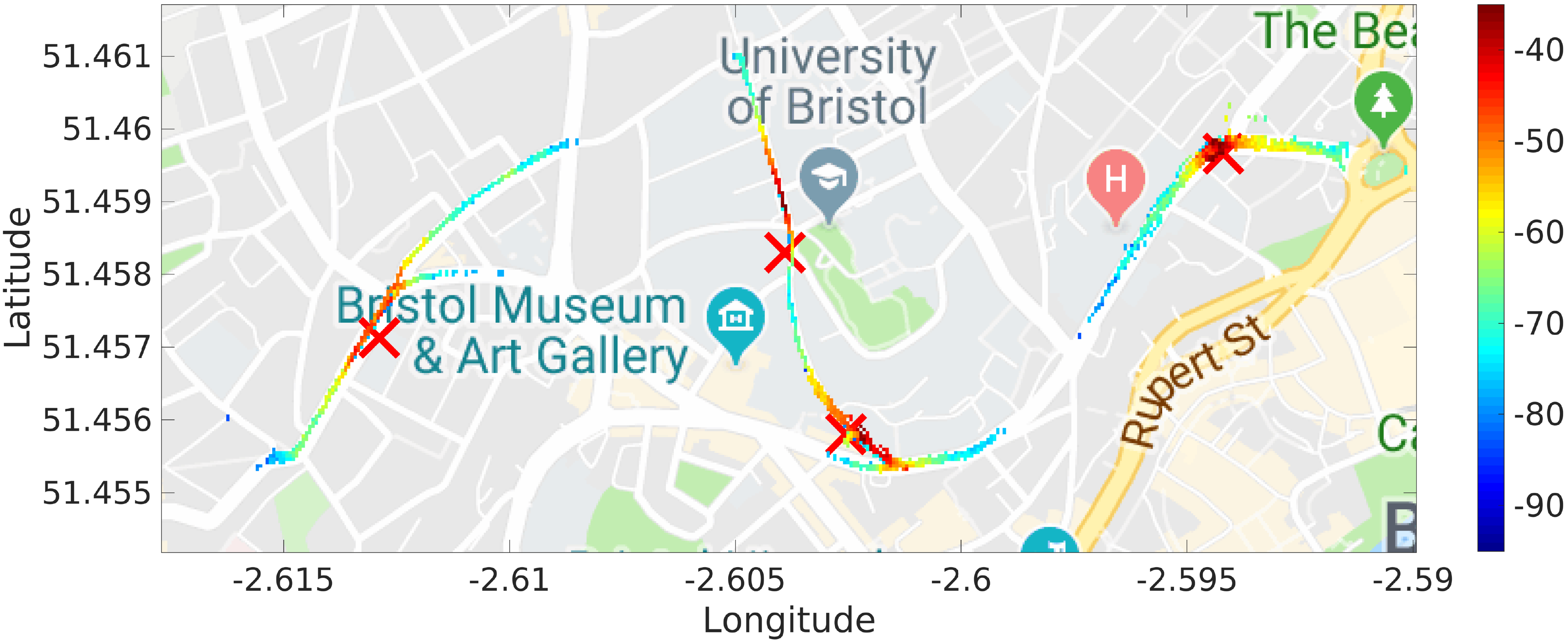}
}
\\
\subfloat[HP Transceiver]{
    \includegraphics[width=1\columnwidth]{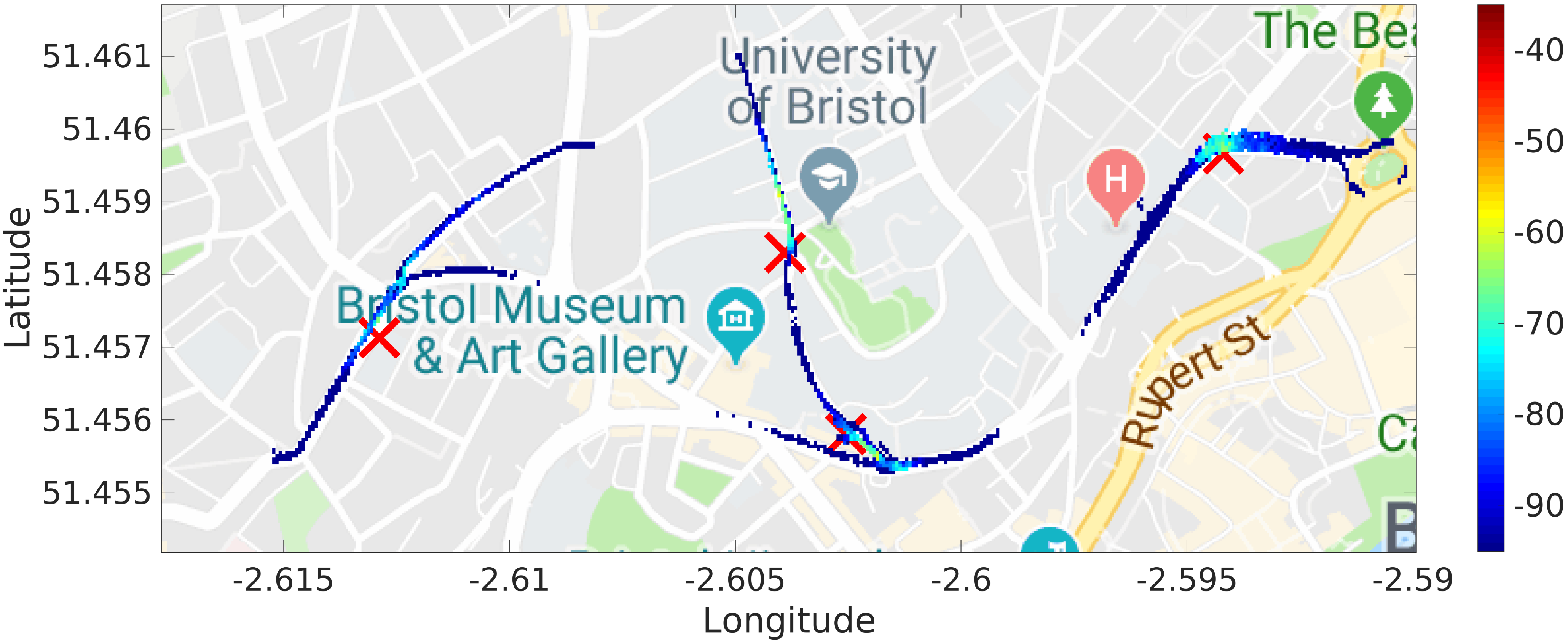}
}
\caption{Heatmaps for the RSSI (expressed in $\SI{}{\dBm}$) associated with the fourth day of trials (ISM band).}
\label{fig.4}
\end{figure}

\begin{figure}[t]
\centering
\subfloat[LP Transceiver]{
    \includegraphics[width=1\columnwidth]{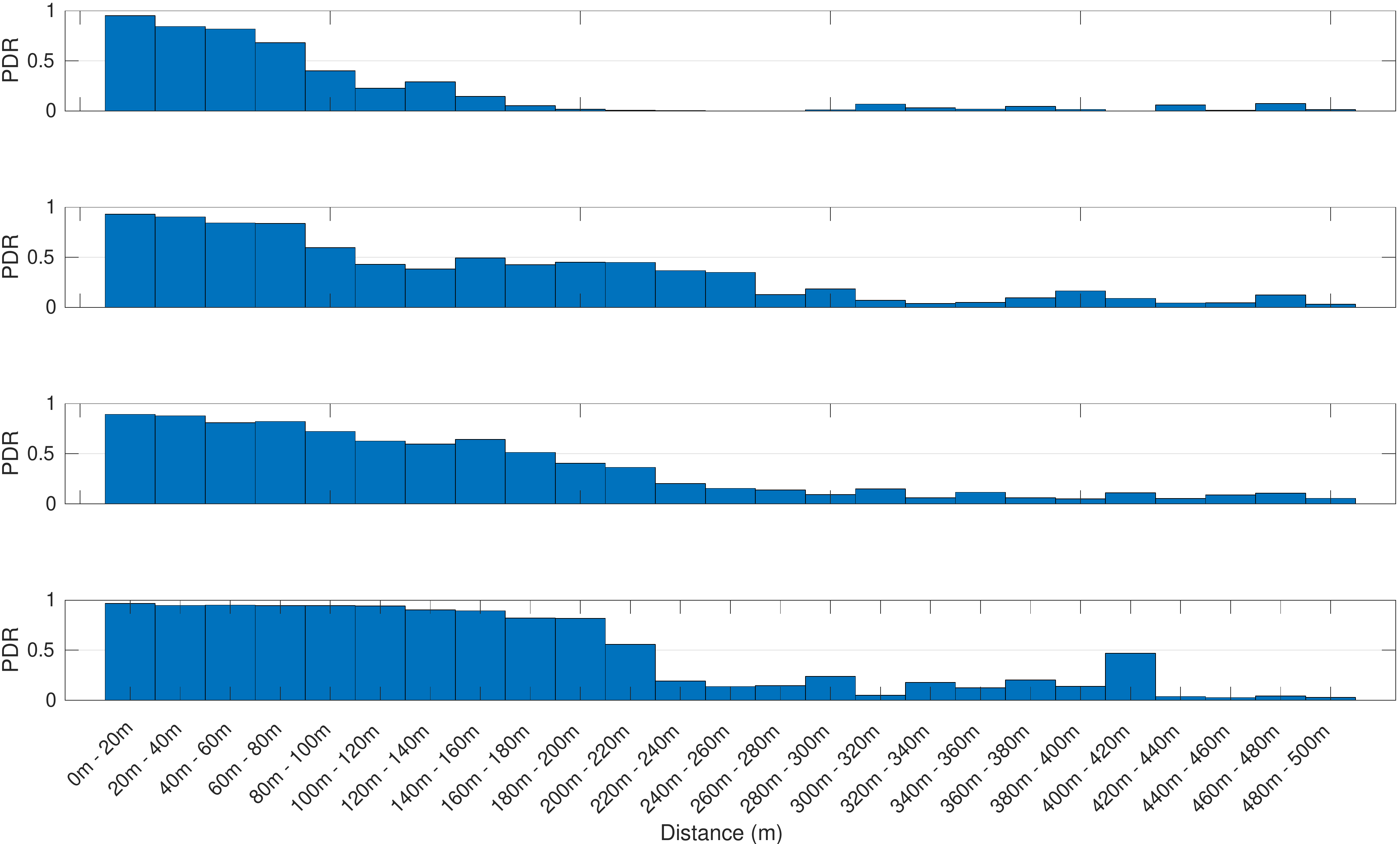}
}
\\
\subfloat[HP Transceiver]{
    \includegraphics[width=1\columnwidth]{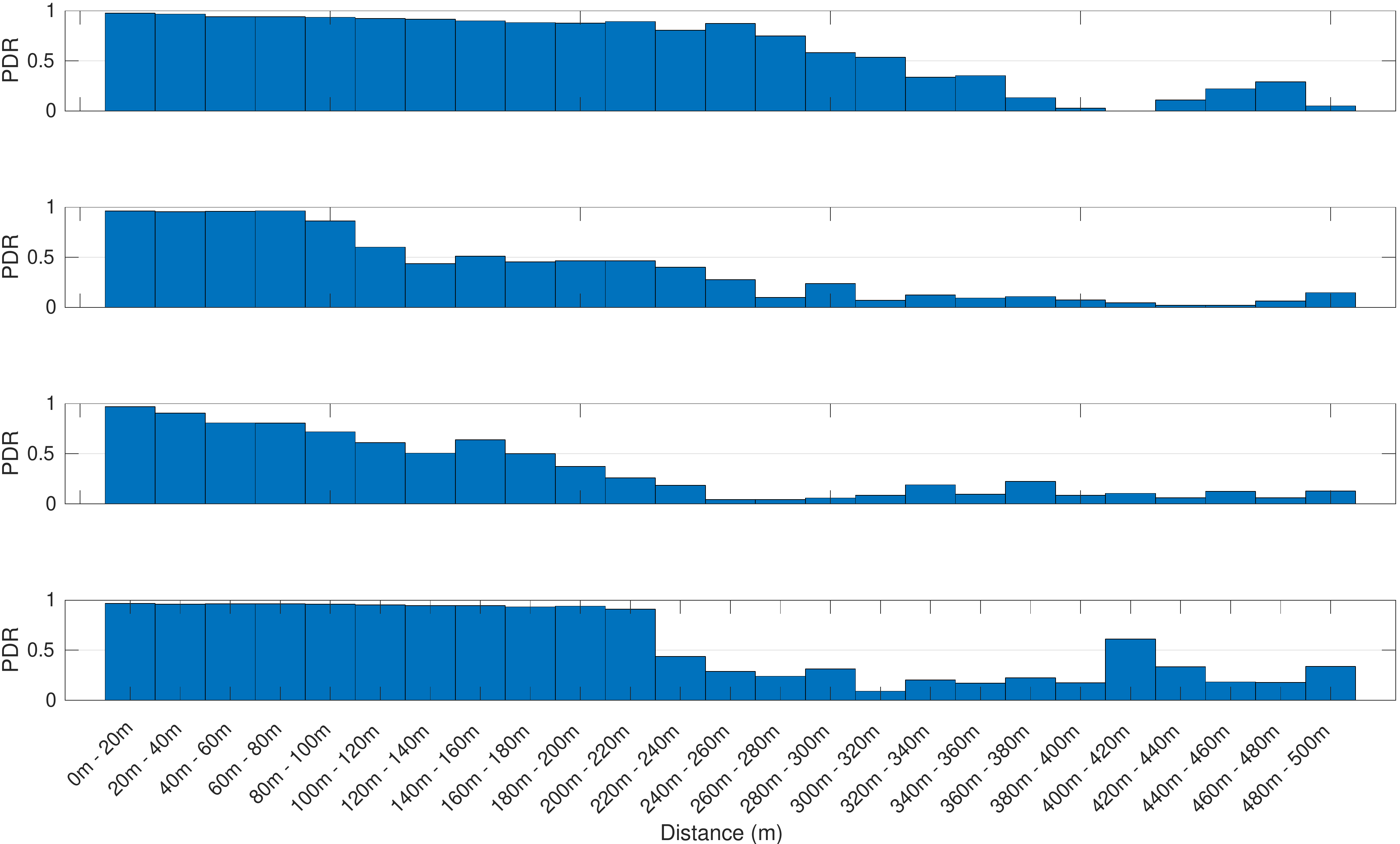}
}
\caption{Awareness horizon associated with the first day of trials (DSRC band) and the OBU-00. Each group of four figures shows the awareness horizon associated with the RSU at sites MVB, DH, HW and SU.}
\label{fig.5}
\end{figure}

\begin{figure}[t]
\centering
\subfloat[LP Transceiver]{
    \includegraphics[width=1\columnwidth]{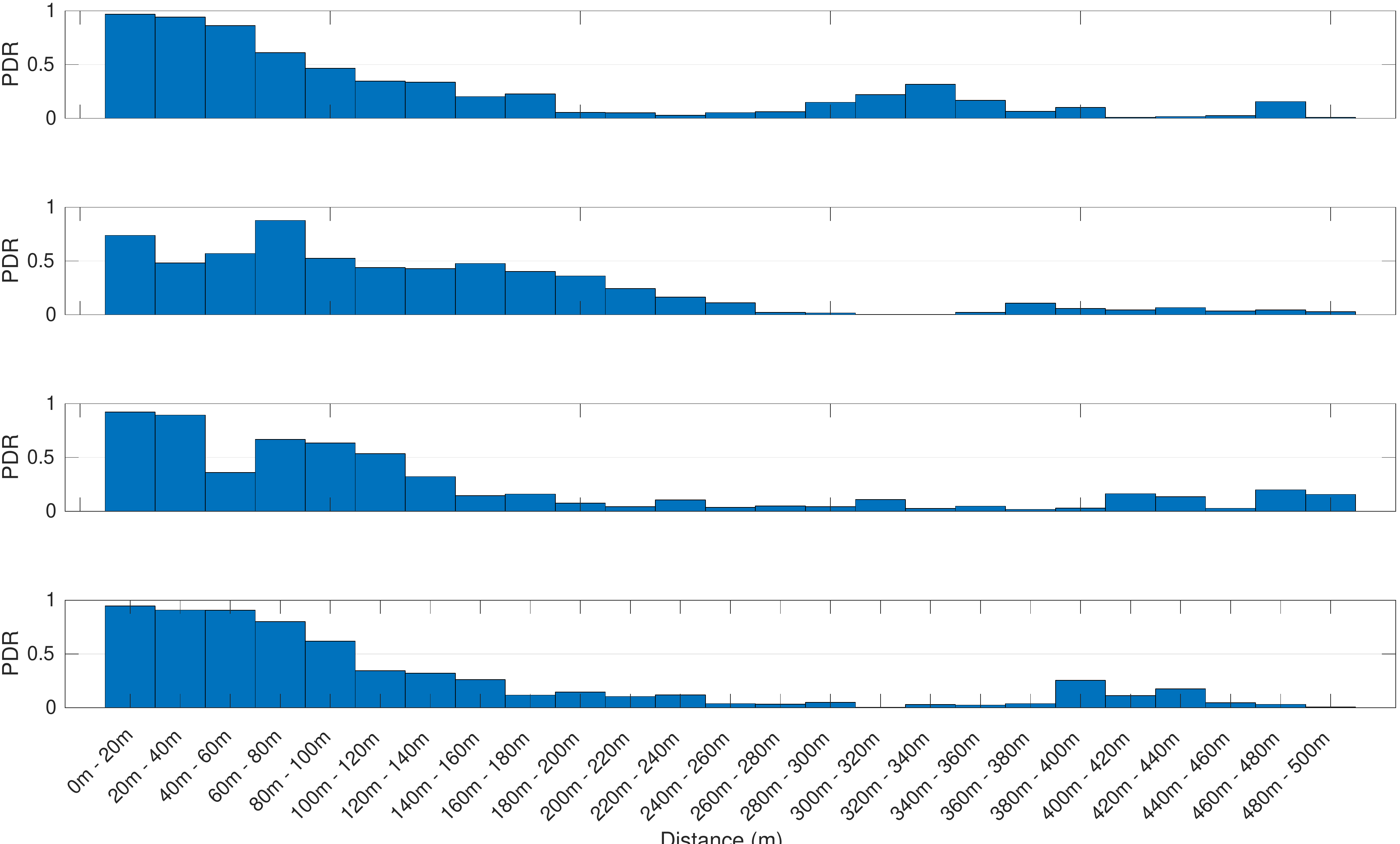}
}
\\
\subfloat[HP Transceiver]{
    \includegraphics[width=1\columnwidth]{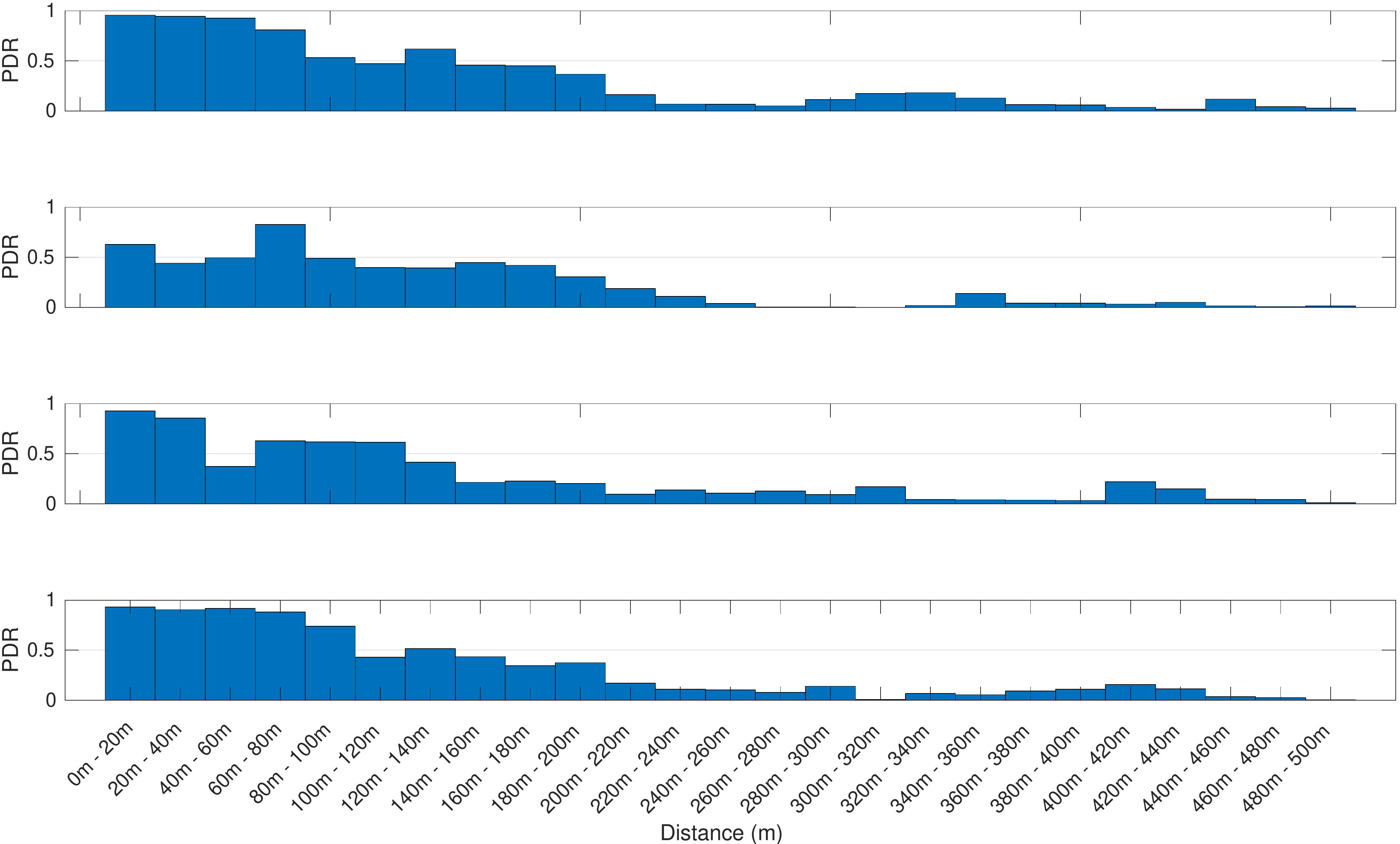}
}
\caption{Awareness horizon associated with the third day of trials (ISM band) and the OBU-00. Each group of four figures shows the awareness horizon associated with the RSU at sites MVB, DH, HW and SU.}
\label{fig.6}
\end{figure}

\begin{figure}[t]
\centering
\subfloat[LP Transceiver]{
    \includegraphics[width=1\columnwidth]{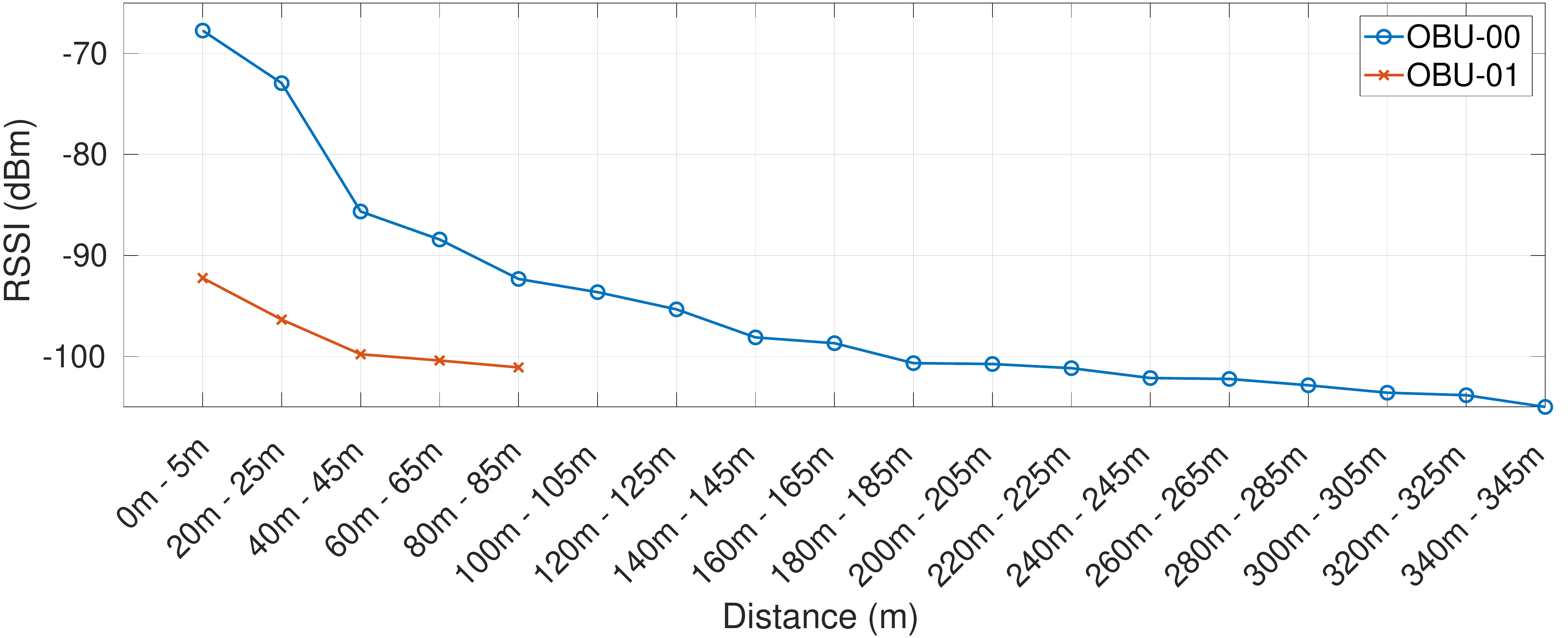}
}
\\
\subfloat[HP Transceiver]{
    \includegraphics[width=1\columnwidth]{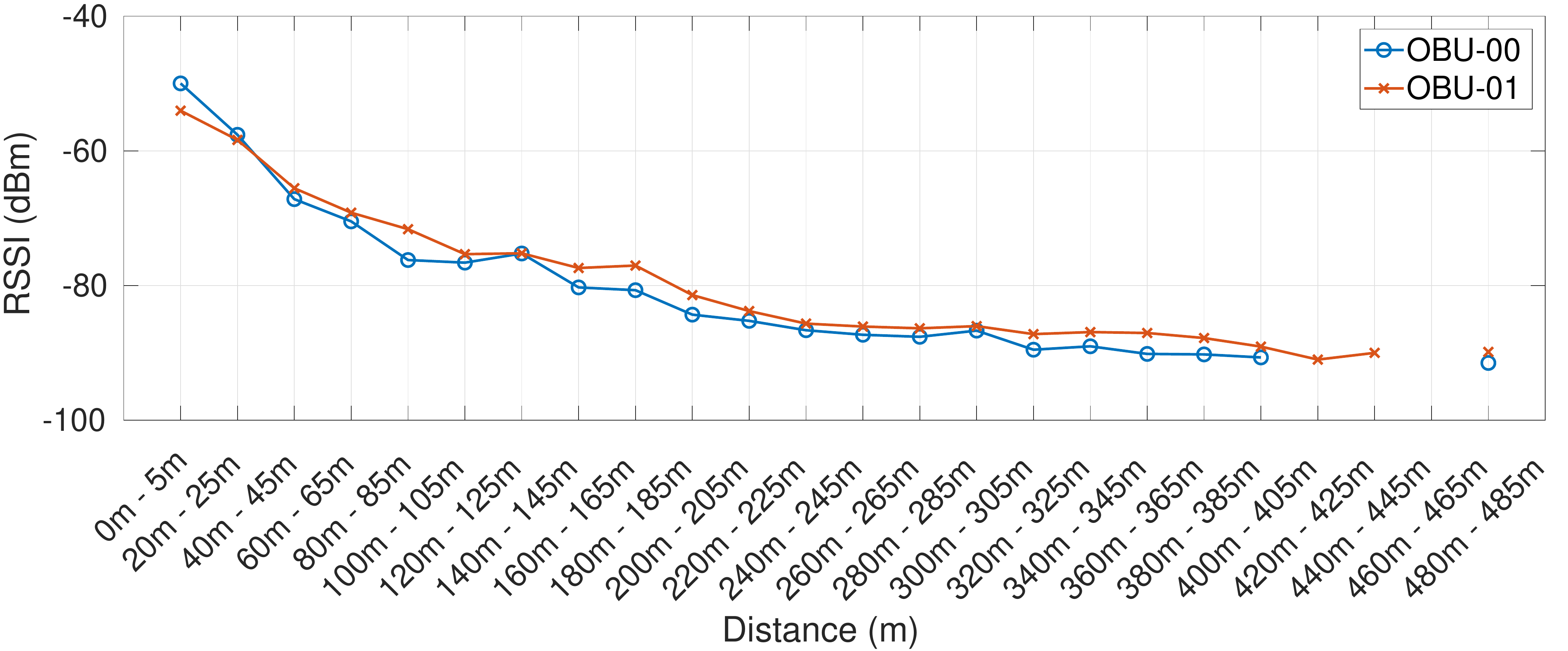}
}
\caption{RSSI measurement as a function of the distance, associated with the first day of trials (DSRC band) and the RSU at the DH site.}
\label{fig.7}
\end{figure}

\begin{figure}[t]
\centering
\subfloat[LP Transceiver]{
    \includegraphics[width=1\columnwidth]{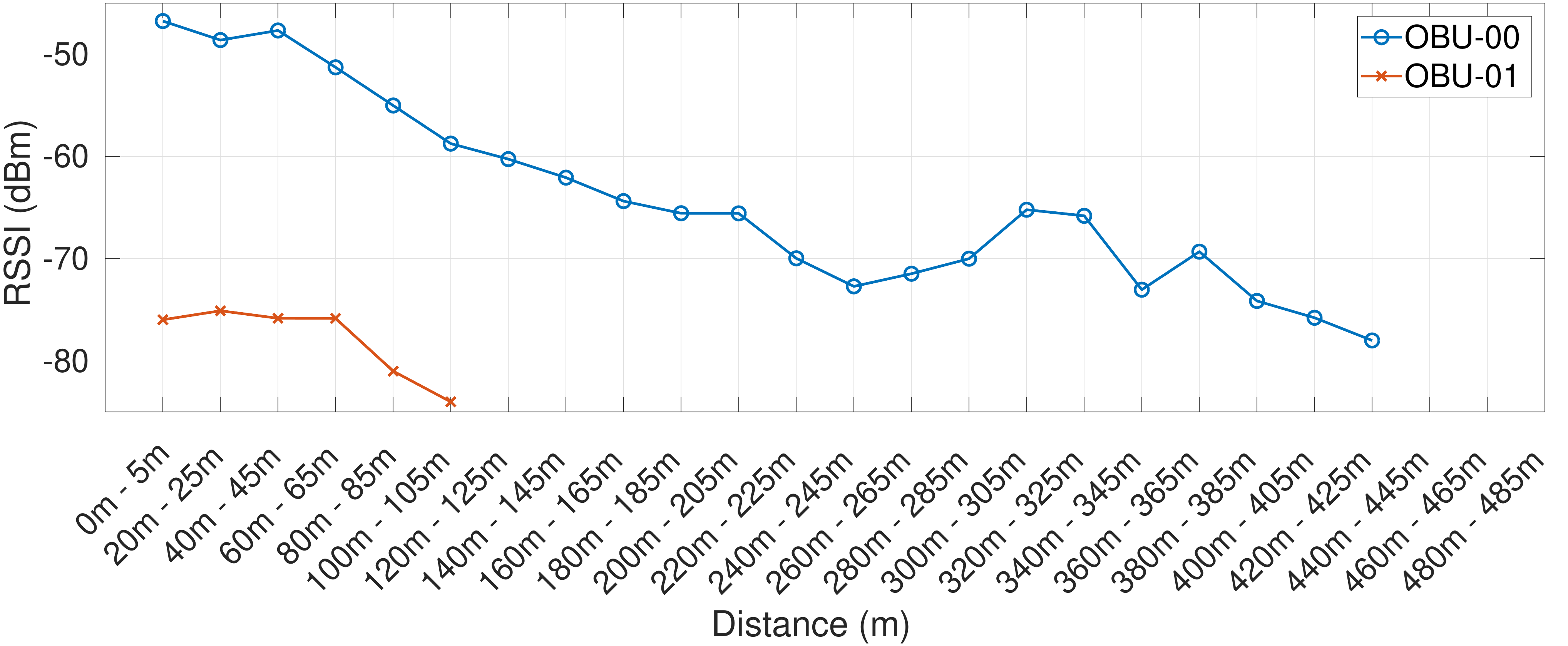}
}
\\
\subfloat[HP Transceiver]{
    \includegraphics[width=1\columnwidth]{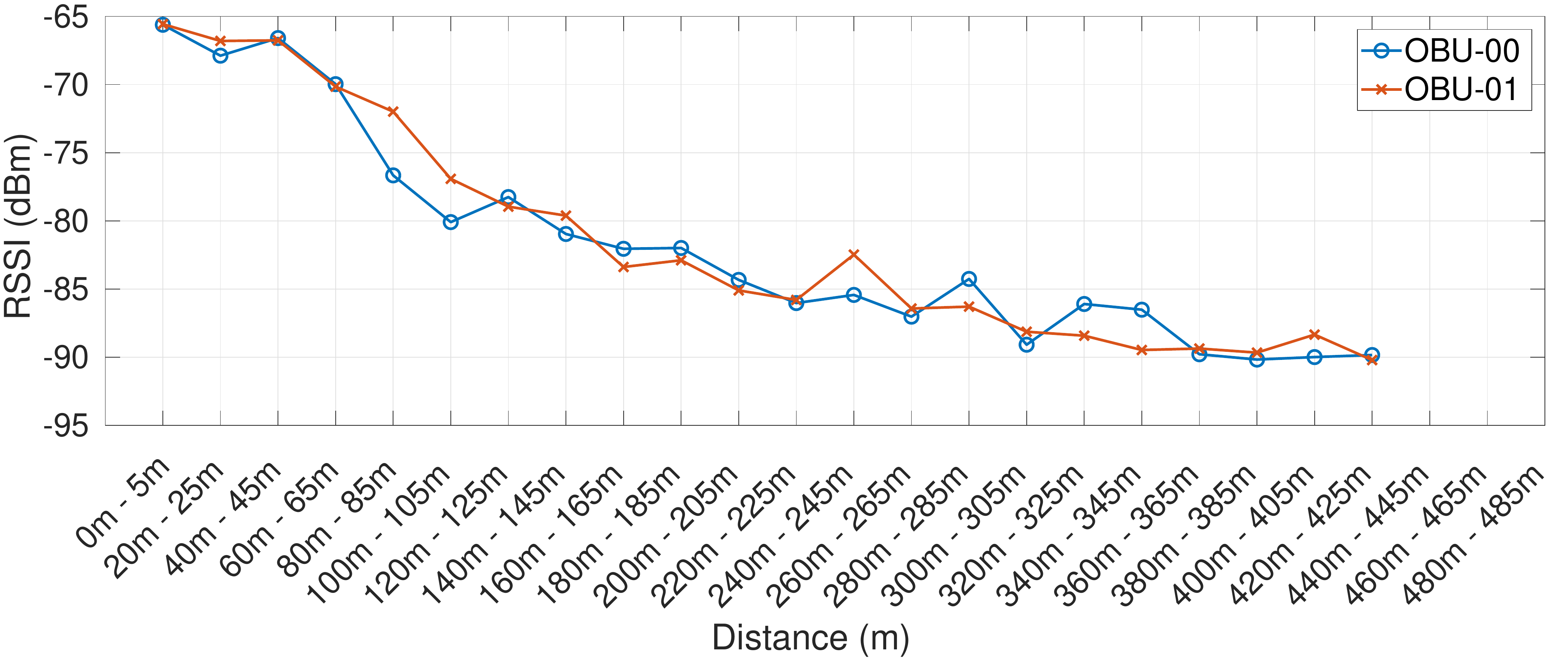}
}
\caption{RSSI measurement as a function of the distance, associated with the fourth day of trials (ISM band) and the RSU at the HW site.}
\label{fig.8}
\end{figure}

\vspace{\baselineskip}
\setlength{\parskip}{0.0pt}
\textbf{Acknowledgments}\par
\setlength{\parskip}{9.96pt}
This work is part of the FLOURISH Project, which is supported by Innovate UK, under Grant 102582.

\bibliographystyle{IEEEtran}
\bibliography{IEEEabrv,bib}
\end{document}